\RequirePackage[hyphens]{url}  
\documentclass[fleqn,10pt]{wlscirep}
\usepackage[utf8]{inputenc}
\usepackage[T1]{fontenc}
\usepackage{bm}
\usepackage{siunitx}
\usepackage{algorithm}
\usepackage{algpseudocode}

\title{Reusable specimen-level inference in computational pathology}

\author[1,2,3,*]{Jakub R. Kaczmarzyk}
\author[1,4]{Rishul Sharma}
\author[1,2]{Peter K. Koo}
\author[1]{Joel H. Saltz}
\affil[1]{Department of Biomedical Informatics, Stony Brook University, Stony Brook, NY, USA}
\affil[2]{Simons Center for Quantitative Biology, Cold Spring Harbor Laboratory, Cold Spring Harbor, NY, USA}
\affil[3]{Medical Scientist Training Program, Stony Brook University, Stony Brook, NY, USA}
\affil[4]{Jericho High School, Jericho, NY, USA}
\affil[*]{jakub.kaczmarzyk@stonybrookmedicine.edu}

\begin{abstract}
Foundation models for computational pathology have shown great promise for specimen-level tasks and are increasingly accessible to researchers. However, specimen-level models built on these foundation models remain largely unavailable, hindering their broader utility and impact. To address this gap, we developed SpinPath, a toolkit designed to democratize specimen-level deep learning by providing a zoo of pretrained specimen-level models, a Python-based inference engine, and a JavaScript-based inference platform. We demonstrate the utility of SpinPath in metastasis detection tasks across nine foundation models. SpinPath may foster reproducibility, simplify experimentation, and accelerate the adoption of specimen-level deep learning in computational pathology research.
\end{abstract}

\begin{document}

\flushbottom
\maketitle

\thispagestyle{empty}

\section*{Main}
\addcontentsline{toc}{section}{Main}

The adoption of deep learning in computational pathology has revolutionized specimen-level analysis, enabling advancements in tasks such as cancer subtyping, metastasis detection, tumor grading, mutation detection, survival prediction, and treatment response evaluation.\cite{echle2021deep, van2021deep, song2023artificialcpath} However, despite the growing application of specimen-level deep learning models, their reusability and accessibility remain significant barriers to widespread adoption.\cite{wagner2022make, wagner2024built} A major contributor to this issue may be the absence of standardized frameworks and tools for sharing and reusing these models.

Recent progress in foundation models for pathology, often hosted on platforms like Hugging Face, has demonstrated the potential of open access to transform the field. These models, designed for general-purpose feature extraction, are readily reusable and have achieved remarkable performance across diverse pathology tasks. \cite{nechaev2024hibou, ai2024kaiko, hoptimus0, filiot2023phikon, filiot2024phikonv2, xu2024gigapath, chen2024uni, vorontsov2024virchow, zimmermann2024virchow2} However, the full potential of foundation models may only be realized when downstream specimen-level models --- those fine-tuned for specific clinical tasks --- are equally accessible. Unfortunately, such downstream models are rarely shared or standardized, creating a bottleneck in the workflow.\cite{wagner2022make, wagner2024built}

Existing toolkits like CLAM \cite{lu2021clam}, Marugoto\footnote{\url{https://github.com/KatherLab/marugoto}}, Slideflow \cite{dolezal2024slideflow}, and STAMP \cite{ElNahhas2024stamp} offer robust pipelines for slide preprocessing and model training. While widely used, these tools primarily focus on model development rather than enabling model reuse. Consequently, researchers may face challenges when attempting to integrate or validate pretrained models across datasets, institutions, or workflows.

\begin{figure}[h]
\centering
\includegraphics[width=\linewidth]{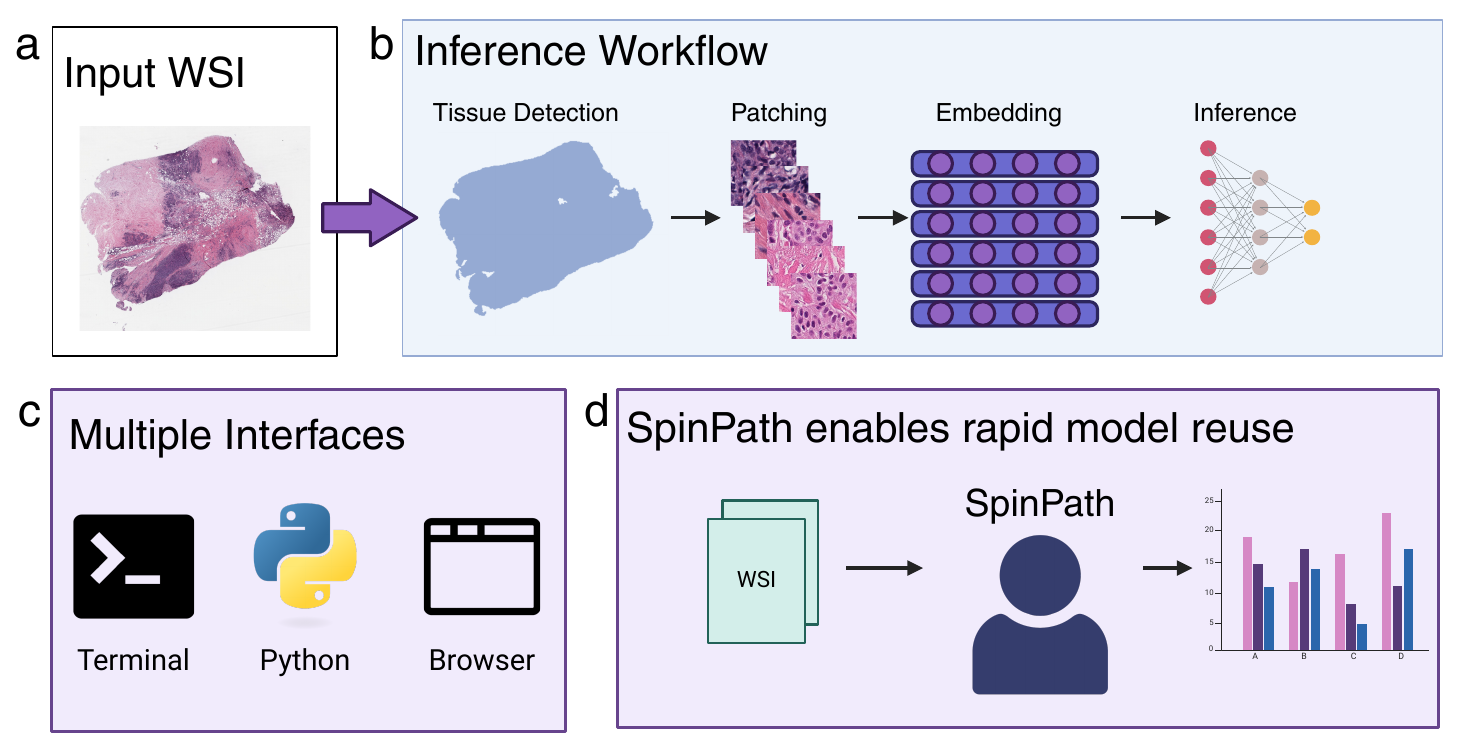}
\caption{\textbf{SpinPath overview.}
The goal of SpinPath is to facilitate the reuse of specimen-level deep learning models in computational pathology.
\textbf{a}, the user first selects a whole slide image (WSI) on which to perform inference.
\textbf{b}, The SpinPath inference workflow follows typical specimen-level inference: tissue detection, extraction of patches, the embedding of patches using a pretrained model, and finally specimen-level inference using a pretrained aggregator model.
\textbf{c}, SpinPath may be used through multiple interfaces, in order to satisfy the needs of users with varying levels of technical experience. 
\textbf{d}, SpinPath assists pathology researchers by enabling rapid reuse of specimen-level models.
Created in \url{https://BioRender.com}.
}
\label{fig:schematic}
\end{figure}

To address these challenges, we aim to democratize specimen-level inference by developing tools that simplify model reuse and foster collaboration. Our solution comprises a collection of software tools, including a Python library, a command-line utility, and a browser-based JavaScript inference tool. These tools, collectively named SpinPath, are designed to streamline the reuse of specimen-level models, facilitate rapid experimentation, support multi-site validation, and enable seamless integration into diverse computational pathology workflows (Fig. \ref{fig:schematic}). In addition, we curated a collection of pretrained specimen-level models, with the goal that the community may contribute additional models. In the present report, we introduce these resources and highlight their potential to bridge the gap between model development and real-world application.

\section*{SpinPath accelerates prototyping and experimentation}

\begin{figure}[h]
\centering
\includegraphics[width=\linewidth]{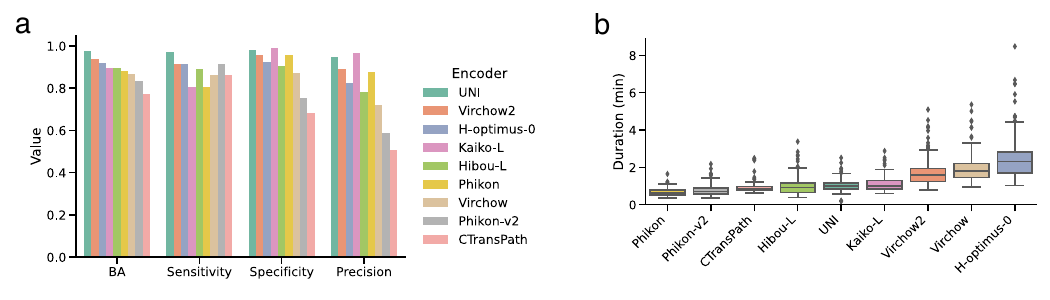}
\caption{\textbf{SpinPath facilitates experimentation across foundation models.}
SpinPath was used to evaluate the performance of metastasis detection. Models were trained on CAMELYON16 \cite{bejnordi2017camelyon16} and applied to SLN-Breast\cite{campanella2019clinical}. All models were accessed from the SpinPath Model Zoo and were downloaded automatically as part of the specimen-level inference workflow.
\textbf{a}, Bar plot of classification metrics on SLN-Breast dataset.
\textbf{b}, Box plot of the durations of inference, grouped by foundation model used to embed tiles. Box plots show the first and third quartiles, the median (central line), and the range of data, with outliers shown as points above and below whiskers.
(BA: balanced accuracy.)
}
\label{fig:sln_breast_results}
\end{figure}

To demonstrate the potential of SpinPath to simplify and democratize specimen-level deep learning analyses in computational pathology, we applied pretrained metastasis detection models in the SpinPath model zoo to an external dataset, SLN-Breast.\cite{campanella2019clinical} The SLN-Breast dataset, distinct from the CAMELYON16 dataset on which the models were trained, highlights the importance of comparing model performance across datasets to understand their generalizability. The UNI model achieved the highest balanced accuracy (BA) of 0.975, followed by Virchow2 (BA=0.937) and H-optimus-0 (BA=0.921). The other models (other than CTransPath) achieved BA above 0.83 but had varying trade-offs among sensitivity, specificity, and precision. Kaiko-L, for instance, achieved a BA of 0.897, with a near-perfect specificity of 0.989 but sensitivity of 0.806 (Fig. \ref{fig:sln_breast_results}a). The variations among models underscore how SpinPath enables users to identify the most suitable model for their specific dataset and task. Moreover, by reducing the technical barriers to deploying and evaluating state-of-the-art deep learning models, SpinPath may accelerate experimental workflows and foster reproducibility, making it a valuable tool for the computational pathology community.

In addition to performance metrics, the runtime of each model during inference was measured to evaluate computational efficiency. The fastest model was Phikon, taking an average of 40 seconds per WSI. The UNI model, which performed highest in metastasis classification on SLN-Breaset, took 63 seconds per WSI on average. H-optimus-0 had the longest average runtime at 154 seconds per WSI (Fig. \ref{fig:sln_breast_results}b). These results highlight the variability in computational demands across foundation models and underscore the efficiency of SpinPath in enabling quick evaluations even with resource-intensive models.

\section*{Code-free, serverless inference in the browser}

The SpinPath JavaScript-based browser tool within SpinPath performs specimen-level inference with no requirement for coding or environment setup. It runs inference entirely in a browser and does not use a server, which may enhance privacy. To demonstrate its usefulness, we used the SpinPath browser tool to detect metastases in WSIs in SLN-Breast. Both specimens were successfully detected as positive for metastasis, and this processing was accomplished in less than one minute per slide. This browser-based approach eliminates common barriers to entry, such as the need for computational infrastructure or programming expertise. It is well-suited for educational purposes, enabling students and researchers to interact with state-of-the-art models without requiring local installations or computational resources. Additionally, it provides a convenient platform for small-scale experiments and quick evaluations of models on new slides, such as assessing cross-dataset generalization. Despite its current speed limitations, this proof-of-concept tool demonstrates the feasibility of serverless, client-side specimen-level inference in computational pathology. Future optimization efforts could significantly enhance its runtime and expand its applicability.

\section*{Methods}

\subsection*{SpinPath Python package}

The SpinPath Python package includes an inference engine, API to interact with the model zoo, and a command line interface (described below). The main purpose of this package is to perform specimen-level inference. To perform inference, the user must provide a pretrained model and a whole slide image (WSI). Once given, SpinPath takes the following steps. First, the tissue regions are detected, and patch coordinates are calculated in the tissue. Next, the patch embedding model is loaded (downloaded from Hugging Face Hub if necessary), and is applied to each patch. The patches are loaded lazily at embedding time. Next, the pretrained specimen-level model is loaded (and is downloaded from Hugging Face Hub if necessary). Then, the embeddings are fed to the specimen-level model, and finally the model results are returned. The SpinPath Python package uses TiffSlide \cite{poehlmann2022tiffslide} to load WSIs and PyTorch to perform deep learning inference. The package may be used as a Python package or as a command line tool.

\subsection*{SpinPath command line tool}

SpinPath offers users a command line tool with which to perform specimen-level inference. This tool is implemented in Python. The tool requires two inputs from the user at minimum: a WSI file and a pretrained specimen-level model. This pretrained specimen-level model is the name of a Hugging Face Hub repository that contains the pretrained model as well as a configuration. The configuration specifies data about how to use the model, including the name of the patch encoder, the size of the patches, and the names of the model outputs. The command line tool is implemented using Click.

\subsection*{SpinPath JavaScript tool}

The SpinPath JavaScript tool allows the user to perform specimen-level interface with no requirement for coding or environment set up. The tool is fully customizable, allowing the user to select their preferred feature extractor and image analysis model. The Vue.js framework is then used to dynamically change which model and feature extractor is used by the tool and display model and feature extractor information.

To start using the SpinPath JavaScript tool, the user first selects a WSI and uploads it to the program. The OpenSeadragon.js package is then used to display the image in a viewer, allowing the user to move and zoom into the image as needed. Once loaded, the user can then select regions for analysis using the Annotorious package. The user can then click the “Run Model Analysis” button to run their selected model on the annotated regions. Using the GeoTiff.js and Transformers.js libraries to access the WSI data and implement the user’s selected model and feature extractor in JavaScript, respectively, The program then automatically handles patching, patch embedding, and aggregating results. The results and logits are then displayed to the user, and the user is given the option to download the patch coordinates and attention scores in the form of GeoJSON data.

\subsection*{Model zoo development}

Models are hosted on Hugging Face Hub in individual, self-contained repositories. These models may be used on their own without SpinPath. The code to train each model is included in each repository. All specimen-level models were trained using NVIDIA GeForce RTX 2080 Ti GPUs.

\section*{Data availability}
The CAMELYON16 dataset may be accessed at \url{https://camelyon16.grand-challenge.org/}. The SLN-Breast dataset may be accessed at \url{https://doi.org/10.7937/tcia.2019.3xbn2jcc}.

\section*{Code availability}
The SpinPath Python toolkit is available at \url{https://github.com/SBU-BMI/SpinPath}. The SpinPath JavaScript tool will be made available following publication.

\section*{Acknowledgements}
We would like to acknowledge the Department of Biomedical Informatics at Stony Brook University and the Simons Center for Quantitative Biology at Cold Spring Harbor Laboratory. JRK would also like to acknowledge the Medical Scientist Training Program at Stony Brook University and NIH grant T32GM008444 (NIGMS).

\section*{Author contributions}
JRK conceptualized the project, developed the Python-based toolkit, prepared datasets, trained specimen-level models, curated the model repository, and conducted experiments. RS and JRK developed the JavaScript-based tool. JHS and PKK provided supervision and guidance throughout the project. JRK and RS wrote the manuscript, and all authors provided revisions. All authors reviewed and approved the final manuscript.

\section*{Competing interests}
The authors declare the following competing interests: J.H.S. is co-founder and chief executive officer of Chilean Wool, LLC. All other authors declare no competing interests.

\bibliography{main}

\end{document}